# Origin and nature of killer defects in 3C-SiC for power electronic applications by a multiscale atomistic approach

Emilio Scalise,[a] Luca Barbisan,[a] Andrey Sarikov,[a,b] Francesco Montalenti,[a] Leo Miglio,[a] and Anna Marzegalli *[a,c]

3C-SiC epitaxially grown on Si displays a large wealth of extended defects. In particular, single, double and triple stacking faults (SFs) are observed in several experiments to coexist. Overabundance of defects has so far limited the exploitation of 3C-SiC/Si for power electronics, in spite of its several ideal properties (mainly in terms of wide gap, high breakdown fields and thermal properties) and the possibility of a direct integration in the Si technology. Here we use a multiscale approach, based on both classical molecular dynamics (MD) simulations and first-principle calculations, to investigate in-depth the origin, nature and properties of most common 3C-SiC/Si(001) extended defects. Our MD simulations reveal a natural path for the formation of partial dislocation complexes terminating both double and triple SFs. MD results are used as input for superior DFT calculations, allowing us to better determine the core structure and to investigate electronic properties. It turns out that the partial dislocation complexes terminating double and triple SFs are responsible for the introduction of electronic states significantly filling the band gap. On the other hand, individual partial dislocations terminating single SFs only induce states very close to the gap edge. We conclude that partial dislocation complexes, in particular the most abundant triple ones, are killer defects in terms of favoring leakage currents. Suggestions coming from theory/simulations for devising a strategy to lower their occurrence are discussed.

a. L-NESS and Department of Materials Science, Università degli Studi di Milano-Bicocca, via Cozzi 55, I-20125, Milano, Italy.

b. V. Lashkarev Institute of Semiconductor Physics, National Academy of Sciences of Ukraine, 45 Nauki avenue, 03028 Kyiv, Ukraine.

c. L-NESS and Department of Physics, Politecnico di Milano, via Anzani 42, I-22100, Como, Italy

## Introduction

SiC is a group IV compound with unique physical and chemical properties, mainly conditioned by the strong tetrahedral bonding between Si and C atoms (4.6 eV). It is a wide bandgap material (2.3-3.3 eV, depending on the crystal phases), which possesses excellent thermal stability (sublimation temperatures up to 2800°C), high thermal conductivity (3.2-4.9 W/cm K, about three times higher than the one of Si), high breakdown field ($30\times10^5$ V/cm, about ten times higher than that for Si), large carrier saturation velocity ($2\times10^7$ cm/s), hardness (Knoop hardness of 2480 kg/mm$^2$), and chemical inertness.[1-3] In fact, the hexagonal phases of SiC are increasingly used for electronic devices operating at elevated temperatures, large power and high frequencies, as well as in harsh environments.[4] Beyond the tetrahedral coordination at first neighbours, silicon and carbon atoms in SiC may occupy different lattice positions at further neighbours, giving rise to multiple crystal structures.[5-7] In particular, the most common SiC polytypes are the hexagonal 6H and 4H phases, and the cubic 3C structure, with a stability hierarchy depending on temperature, as recently understood in our recent paper.[7] The 3C-SiC cubic phase is the natural output of deposition processes on Si substrates and displays several interesting features. First, it has excellent electrical properties, such as a reduced phonon scattering resulting from the high symmetry of this material, as well as the highest electron mobility and saturated drift velocity, when compared to other SiC polytypes. Second, 3C-SiC may be used as a substrate for the epitaxial growth of relevant materials, such as gallium nitride[8] and boron nitride.[9]

Currently, the growth of large area bulk 3C-SiC material is still under development. 3C-SiC epilayers can be deposited on hexagonal SiC substrates (4H and 6H). Unfortunately, commercially available wafers of hexagonal SiC are small, much smaller than the Si ones, and their crystal quality is not sufficiently good as well. They also contain a large number of defects that may propagate in the 3C-SiC epilayers.[10] Alternatively, cubic silicon carbide layers can be heteroepitaxially grown on (001) and (111) silicon substrates, enabling direct integration into the Si technology and thus offering an advantage over other SiC polytypes.[11] Still, structural and thermal differences between the substrate and epilayer hinder the possibility to obtain high-quality crystalline 3C-SiC on Si substrates, thus reducing the high theoretical potential of this material for power electronics. In particular, the lattice parameters of cubic silicon carbide and silicon (4.53 Å and 5.43 Å, respectively, at room temperature) differ by about 20%, and their thermal expansion coefficients by 8%. Thus, both during the high temperature deposition and the cooling down, several types of defects, such as dislocations, twins and stacking faults nucleate at or close to the 3C-SiC/Si interface and propagate into the 3C-SiC films, as driven by the misfit strain relaxation.[12,13] Such defects have detrimental effects on the performance of devices, particularly on the leakage current and breakdown electric field. Their correlation with twins and stacking faults in the epitaxial 3C-SiC films has been observed experimentally in a number of publications.[14-21] As a consequence, although the Si substrates are considered at the moment as the most convenient choice for 3C-SiC epitaxy, the high-quality growth is still an important challenge for researchers.

A common strategy to reduce the 3C-SiC and Si lattice misfit strain at the early stage of the Si carbide epitaxy is the so-called carbonization process, realized by a deposition of carbon-rich precursors under hydrogen flux.[22,23] Release of strain and associated stress takes place by different mechanisms, depending on the specific carbonization procedure and the substrate orientation.[24-26] As a result of this procedure, only some residual strain remains in the layer, with sign and magnitude depending on the deposition parameters.[25] In most cases, an array of ordered misfit sessile dislocations appears at the 3C-SiC/Si substrate interface, allowing to match five SiC unit cells to four Si lattice units,[26,27] and leaving a slightly compressed 3C-SiC layer (at the growth temperature) for the subsequent deposition.[28] The carbonization step does improve the crystal quality of 3C-SiC layers, but it does not reduce the defect density to such an extent to ensure a quality level (i.e. staking fault density to $10^2$/cm) acceptable for the devices.[17,27,29]

In addition to sessile dislocation arrays formed at the 3C-SiC/Si interface, the stress in the epitaxial layer is partially released by nucleation and evolution of the partial dislocations accompanying the appearance of single and multiple stacking faults (SFs) in the 3C-SiC films, as favoured by SiC polytypism. In fact, hexagonal phases become favoured at high deposition temperatures (T=1200-1400°C) and inclusion of hexagonal polytypes in 3C-SiC effectively corresponds to the multi-plane stacking faults in the cubic lattice.[7,30] These multiple stacking faults in cubic silicon carbide are predicted to have lower formation energies at any temperature, as compared to the single-plane SF,[7] and indeed their abundance in 3C-SiC epitaxial layers is confirmed experimentally.[26,27,31,32]

The observed direct correlation between the stacking fault density and the leakage current values in 3C-SiC layers,[14-21] discussed above, drove several difficult attempts in finding the way for their reduction.[18,21,33,34,35,36] Actually, it is an open question whether the SFs per se are responsible for the detrimental leakage currents, because they are theoretically predicted to introduce no mid-gap electronic states in 3C-SiC phase.[37,38] Instead, partial dislocations and dislocation complexes, bounding the stacking faulted regions to the rest of the 3C-SiC crystal, may be well considered possible candidates as killer defects in 3C-SiC, because inducing leakage currents.

We use multiscale approach combining classical molecular dynamics (MD) simulations and ab initio calculations, to study the formation mechanisms and the electronic properties of complexes of partial dislocations in 3C-SiC epitaxial layers. We show that the strain conditions developed in the early deposition stages enhance the probability of obtaining three partial dislocations in adjacent planes, terminating a triple staking fault, in agreement with experimental indications.[39] The attractive interaction between the dislocations induces the formation of stable complexes, which are particularly interesting for the transport properties. In fact, a detailed analysis of the electronic properties of the single partial dislocations, as well as of the dislocation complexes, is performed by ab initio calculations, demonstrating the appearance of defect states in the middle of the band gap of 3C-SiC, particularly in the case of double and triple partial dislocation complexes.

The paper is organized as follows. In **Results and discussion**, we first describe the technical details of the molecular dynamics (MD) simulations and ab initio calculations (subsection **Methods**). Then we present the discussion of obtained results in two parts: results on the origin of the dislocation complexes and their final structure obtained by MD simulations are discussed and then the electronic properties of the partial dislocation complexes terminating single, double and triple stacking faults obtained by ab initio calculations, are shown. Finally, we draw conclusions to the paper.

## Results and discussion

### Methods

**Molecular Dynamics.** The evolution of partial dislocations and the formation of dislocation complexes terminating the stacking faults in 3C-SiC have been simulated by molecular dynamics, employing the Large-scale Atomic/Molecular Massively Parallel Simulator (LAMMPS).[40] The simulations have been performed mainly within the Vashishta potential,[41] which has some advantages over the few empirical potentials fitted on SiC parameters. In particular, this potential takes into account the interactions beyond the first neighbour shell, therefore it provides a different total energy for the hexagonal versus the cubic configuration, thus enabling to compute one empirical estimate of SF energies. Moreover, the comparatively longer times for individual step computation is compensated by the possibility to choose significantly larger time step values, as compared to other potentials (e.g. 1 fs ensuring relative energy conservation to $10^{-5}$). The main results of our simulations have been also verified by additional simulations using the Analytic Bond Order Potential (ABOP),[42] which is complementary to the Vashista potential, according to the analysis outlined in Ref. [43].

The simulations have been performed within the framework of the Nose-Hoover thermostat. Positions and velocities of atoms have been sampled from the canonical (NVT) ensemble and obtained from non-Hamiltonian equations of motion.[44] The atomic trajectories have been analysed using the Open Visualization Tool (OVITO) software,[45] which enables to resolve extended defects in virtual crystals, highlighting SFs and providing dislocation lines with the associated Burgers vectors. The simulation cell has been composed as an orthogonal box defined by $\vec{s} = \frac{a}{2}[1\bar{1}0]$, $\vec{v} = \frac{a}{2}[110]$, $\vec{w} = a[001]$. Periodic boundary conditions have been applied in the $\vec{s}$ and $\vec{v}$ directions to model an infinitely large slab of 3C-SiC. To model the effect of a thick bulk substrate in the 3C-SiC layer, the bottom atoms in the simulation cell have been kept immobile, while the free surface condition (surface atoms exposed to vacuum) has been set on the opposite side along the $\vec{w}$ direction (top part of the cell). Box dimensions have been chosen in order to ensure the optimum balance between the computational cost and the accuracy in the description of dislocation gliding in the considered systems.

Dislocations have been inserted in the simulation cell by shifting each atom according to the displacement field vectors calculated within the dislocation modelling in the framework of the linear elasticity theory.[46] For an arbitrary dislocation, the field components can be easily calculated rotating the reference system and decomposing the dislocation Burgers vector into a screw ($b_{//}$) and an edge component ($b_\perp$). At this, $\hat{z}$ axis has to be oriented along the dislocation line $\vec{\xi}$, $\hat{x}$ axis has to be perpendicular to $\vec{\xi}$ in the glide plane, and $\hat{y}$ axis oriented following the right-hand rule, respectively. In this case, the components of the displacement field vectors can be written as follows:

$$u_x(x,y,z) = \frac{b_\perp}{2\pi}\left(\arctan(\frac{y}{x}) + \frac{xy}{2(1-\nu)(x^2+y^2)}\right) \quad (1)$$

$$u_y(x,y,z) = -\frac{b_\perp}{2\pi}\left(\frac{(1-2\nu)}{4(1-\nu)}\ln(x^2+y^2) + \frac{x^2-y^2}{4(1-\nu)(x^2+y^2)}\right) \quad (2)$$

$$u_z(x,y,z) = \frac{b_{\|}}{2\pi}\arctan\left(\frac{y}{x}\right) \quad (3)$$

where ν = 0.25 is the Poisson ratio of 3C-SiC.[47]

Having a free surface and, in fact, a periodic array of dislocations, some corrections to the displacement field are needed. First, the stress normal to free surface should be zero after system equilibration. In order to save computational time, we have added image dislocations in the vacuum above the free surface.[46] Second, as proved in Ref.[48], the equations above if summed in an infinite series turn to be conditionally convergent, so that a correction to the displacement field in the plane $\hat{x}\hat{y}$ is needed to account for an infinite array of parallel dislocations.[49]

As a first stage, we considered infinitely long and straight dislocation lines with $\vec{\xi}//\vec{v}$ (simulation set called A in the following). The simulation cell was composed of 40320 atoms arranged in the 21.6 x 1.23 x 15.7 nm³ box. The box dimension along the direction of the vector $\vec{s}$ has been chosen so that a specific strain was present in the cell after the insertion of dislocations. In the direction of the vector $\vec{v}$, the box was narrower to reduce the computational cost, being at the same time large enough to allow kink formation and migration during dislocation gliding. We tested the appropriateness of the box dimension in this direction: three times increase of this value did not effect on the motion of dislocations. The number of atomic layers in the direction of the vector $\vec{w}$ was chosen in order to have the dislocation far enough from both the fixed atoms at the bottom part of the simulation cell, and from the free surface in its top part. A series of simulations has been performed varying the number of atomic monolayers and the distances of dislocations from the surface. The best description was obtained for the 144 monolayers in the direction of the vector $\vec{w}$ of simulation cell.

Thermal annealing of the defective systems has been performed at the thermostat temperature of 1400 K, corresponding to the typical experimental temperature for 3C-SiC deposition (in the case of ABOP potential, the temperature has been increased to 2000 K, in order to facilitate the dislocation glide motion). Thermal dilation of the film has been accomplished scaling the box dimensions, where the scaling factor has been calculated for each potential simulating a 3C-SiC bulk cell in the NPT ensemble, using the Nose-Hoover barostat.

**First principles calculations.** The calculations of the electronic structure have been performed by density-functional theory (DFT) with generalized-gradient (GGA) corrected exchange-correlation potentials,[50] as implemented in the Quantum Espresso package.[51] Plane-wave basis set with a kinetic energy cutoff of 80 Ry have been used, with projector augmented wave (PAW) pseudopotentials,[52] and a 1x4x1 Monkhorst-Pack grid for sampling the Brillouin zone of the 3C-SiC supercell, increased to 1x10x1 for the LDOS calculations. The structural relaxations have been performed until the average atomic force is lower than $10^{-4}$ Ry/Bohr, using a conjugate gradient method and including van der Waals interactions within the semiempirical method of Grimme (DFT-D2).[53] For the band structure and the LDOS calculations, meta-GGA exchange-correlation functionals have been exploited in order to improve the bandgap prediction of 3C-SiC. In particular, by using recently developed non-empirical strongly constrained and appropriately normed (SCAN) meta-GGA functionals[54] a bandgap of about 1.9 eV has been obtained as compared to 1.35 eV of the Perdew-Burke-Ernzerhof (PBE) GGA calculations. By comparing the PBE results and the meta-GGA ones, which are much closer to the experimental values, we understand the impact of the bandgap underestimation on the intra-gap states: the larger gap has mainly the effect of further separating the defect states from the conduction band (CB) minimum, but their position in energy with respect to the valence band (VB) is not much changed. Thus, we conclude that the underestimation of the bandgap that one still has by using the SCAN functionals does not affect our prediction on the position of the intra-gap states with respect to the VB.

We modelled the partial dislocations by using periodic boundary conditions within a supercell approach. The super-cell is orthorhombic, with the three axis corresponding to the <-11-2>, <110> and <-111> directions of the 3C-SiC crystal, and it contains 1056 atoms, having sides of 58.64, 6.15 and 30.15 Å, respectively. Pairs of dislocation dipoles have been introduced in the supercells for the 90° and 30° partials, respectively, thus obtaining a supercell with both the Si- and C-core partial dislocations. For the termination of the double and triple SFs, the simulated supercells contain two dislocations dipoles at the opposite sides of the SFs (indicated as configuration A and B in the next). The distance between the dislocation in dipoles (about 27 Å) has been verified to be large enough to rule out any relevant effects, caused by their interactions, on the predictions of the defect energy states, which are the main purpose of these first principle calculations.

The results are discussed in two following subsections. In the first one, a possible formation mechanism of dislocation complexes and multiple stacking faults is proposed and then corroborated by our MD simulations. In the second part, the ab-initio calculations of the electronic properties of the complexes obtained by the MD simulations are presented.

**Molecular dynamics simulation results**

**Defect formation mechanism by MD.** We exploited molecular dynamics simulations to reveal the key steps of the formation of dislocation complexes, which may be considered as candidates for the killer defects in 3C-SiC layers.

In zinc-blend crystals, which include also 3C-SiC, the primary slip system is {111} <110>, i. e. the glissile dislocations nucleate and move in {111} planes, the dislocation lines are mainly directed along <110> directions, and the deformations induced in the crystals are described by Burgers vectors $\vec{b}$ = a/2<011>. A perfect glissile dislocation can dissociate in two partial dislocations with Burgers vectors $\vec{b}$ = a/2<121> (Shockley dislocations) forming 30° and 90° angles with their respective dislocation lines, and a SF between the two cores.

In the early deposition stages of 3C-SiC layers on Si substrates, first a network of dislocations is formed in the SiC side of the 3C-SiC/Si interface, due to the high tensile misfit strain, and stress induced by it. The main part of the stress relaxation is provided by the direct formation of Lomer dislocations at the interface, i.e. sessile perfect dislocations not related to the stacking faults.[26,27] Moreover, formation of glissile Shockley partial dislocations, which introduce SFs in the 3C-SiC layer may also contribute to the stress release.[26,27] Under tensile strain condition, the 90° partials lead the relaxation process and are driven towards the 3C-SiC/Si interface.[27] The residual strain can be further released during subsequent 3C-SiC layer deposition by the motion of the glissile dislocations, already present in the layer, or by the nucleation of new extended defects at the film surface, and propagation of them into the bulk.

Fig. 1 reports the MD simulations results for the simulation set A. A 90° partial dislocation was inserted deep in the simulation cell, the red colour in Fig. 1 corresponding to the atoms forming the core of this dislocation and the orange to the atoms belonging to the SF, respectively. Notice the compressive lobe above the core of the 90° partial, decreasing the initial tensile strain, shown in the volumetric strain maps in the insets of Fig. 1.

The dimension of the simulation cell along the direction of the vector $\vec{s}$ has been chosen so that the array of 90° partial dislocations created by the periodic boundary condition in this direction would cause the slightly compressive in-plane strain (~ 0.45 %) present after the carbonization procedure, corresponding to the relationship of 5 SiC unit cells on 4 Si cells at the typical growth temperature. Under compressive strain condition, 30° partial dislocations lead the stress relaxation moving from the 3C-SiC surface toward the interface with Si substrate.[55] Therefore, to understand the formation process of dislocation complexes, 30° partial dislocations were also inserted in the simulation cell. In particular they are placed at a distance of 2.5 nm from the free surface of the 3C-SiC film, into the glide planes parallel to the one hosting the 90° dislocation. 30° partial dislocations are shown in Fig. 1 with their cores in green colour. During MD simulated annealing (1400 K, 50 ps), 30° dislocation turns to be always expelled to the surface, as reported in the first row of Fig. 1, unless it is inserted in the plane just adjacent to the SF plane hosting the 90° dislocation, as demonstrated in the second row of this figure. In the latter case, the results of simulations show that the 30° partial is driven away from the 3C-SiC layer surface, while the 90° partial is pulled in the opposite direction (Fig. 1, second row). When the two cores of 30° and 90° partial dislocations are at the closest distance, a new defect is created terminating the double-plane SF. This new defect, known as 30° extrinsic partial dislocation, has been experimentally seen both in 3C-SiC as well as other compounds.[56] The observed dislocation dynamics is due to both the strain release process in the film and the attractive elastic interaction between the two dislocations. The latter effect is demonstrated by the volumetric strain colour maps presented in the second row of Fig. 1. Namely, decreasing the distance between the 30° and 90° dislocations, the colour lobes corresponding to the strain of opposite signs increasingly overlap, with partial mutual cancellation. In the experimental case, the simulated mechanism of the formation of 30° extrinsic partial dislocation should be enhanced by the lower formation energy of double SF with respect to the single one, not accurately reproduced by the potential used for MD simulations.

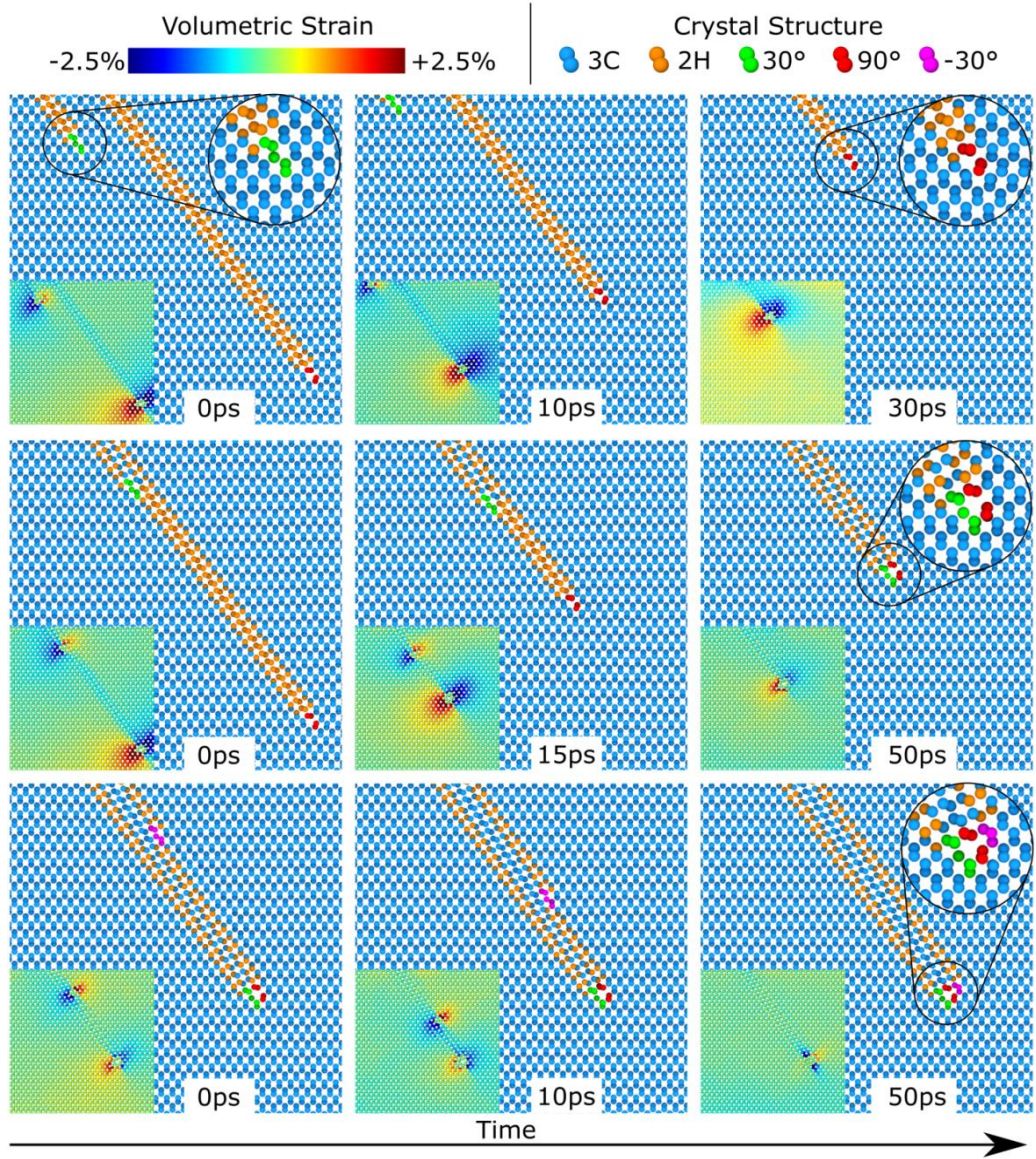

**Fig. 1** MD simulation snapshots of the proposed triple SFs and relative defect termination formation mechanism. First figure row: 30° partial dislocation (core atoms in green) is inserted six {111} planes far from the glide plane of a 90° partial glide plane (core atoms in red). Both dislocations are expelled. Second row: 30° partial dislocation is inserted in the adjacent plane of 90° glide plane. Dislocations move in opposite directions, join and form a 30° extrinsic partial dislocation. Third row: another 30° partial dislocation (core atoms in fuchsia) is inserted in the plane adjacent of the double SF. It reaches the 30° extrinsic partial and forms the stable triple defect. In the insets volumetric strain maps for each snapshot are reported.

Starting from the 30° extrinsic partial dislocation with a double stacking fault, we go on by inserting a second 30° partial dislocation, as illustrated in the third row of Fig. 1, where the core atoms are coloured in fuchsia. This newly inserted dislocation has a screw component of the Burgers vector opposite to the one of already existing 30° partial. Therefore, we label this dislocation as -30° in Fig. 1. The choice of this specific direction of the dislocation Burgers vector has two reasons. First, as 30° partial it releases the residual compressive strain as required by the system. Second, summed with the Burgers vector of already present 30° extrinsic partial, a zero total Burgers vector is resulted, which is the only possible case to form stable triple dislocation complexes, according to our recent findings.[57] Gliding of this dislocation with its SF extension is observed as a result of MD simulations only when it is inserted in the plane adjacent to the double SF. The new dislocation reaches the existing stable defect and forms with it a new defect configuration, which terminates triple SF, composed of three partial dislocations, having the total Burgers vector equal to zero. The motion of the -30° dislocation does not change by inserting it adjacent to the SF formed by 90°, or 30° PD, only being slower in the latter case, due to a weaker driving force. Therefore, the defects with the second type of the dislocation stacking are expected to be less probable to observe experimentally as compared to the first type in defect stacking. Interaction of the 30° partial dislocation with the 30° extrinsic partial dislocation to form a triple dislocation complex takes place according to the mechanism described above for the case of 90° and 30° partial interaction. We note also that the triple SF is equivalent to an inclusion of a hexagonal 6H-SiC stripe and is often defined as a microtwin. The triple dislocation complex terminating the twin inclusion corresponds to the one observed experimentally in phase transitions between 3C and 6H phases,[58] as well as forming the Σ3 {112} incoherent boundaries.[59-61]

**Electronic properties by ab-initio calculations**

**Stacking faults.** The changes in the electronic properties caused by the presence of SFs in 3C-SiC have been investigated by different theoretical groups and there is a general consensus that SFs in 3C-SiC do not give rise to electrical active states in the fundamental bandgap of the material, in contrast to the hexagonal SiC phases.[37,38,62,63] As highlighted above, SFs in 3C-SiC correspond to the inclusions of stripes with hexagonal structure in the cubic phase. The valence band edges of the cubic and hexagonal polytypes are almost aligned,[64] and the difference in the bandgap values between the cubic and the hexagonal phases causes a band-offset mostly between their conduction bands. In particular, the 3C phase has the smallest bandgap as compared to the hexagonal polytypes, thus the inclusion of hexagonal stripes (i.e. SFs) in 3C-SiC creates quantum barriers as evident in Fig. 2, where the local density of states (LDOS) obtained by density functional theory (DFT) calculations has been plotted as a function of the position along the c-axis ([111] direction) of a simulated 3C-SiC supercell, which includes a triple SF (the behaviour with single and double SFs is identical). This is a distinctive aspect of the SFs in 3C-SiC, as compared to the hexagonal polytypes and particularly 4H-SiC, where a quantum well-like behaviour of the SFs has been found,[37] causing the rise of active unoccupied states in the SiC bandgap. On the contrary, SFs in 3C-SiC do not give rise to intra-gap states, which have been inferred as a very likely cause of the leakage current in 3C-SiC devices. However, SFs have finite dimensions in real crystals and their terminations have not yet been included in the analysis of the electronic properties of defective 3C-SiC.

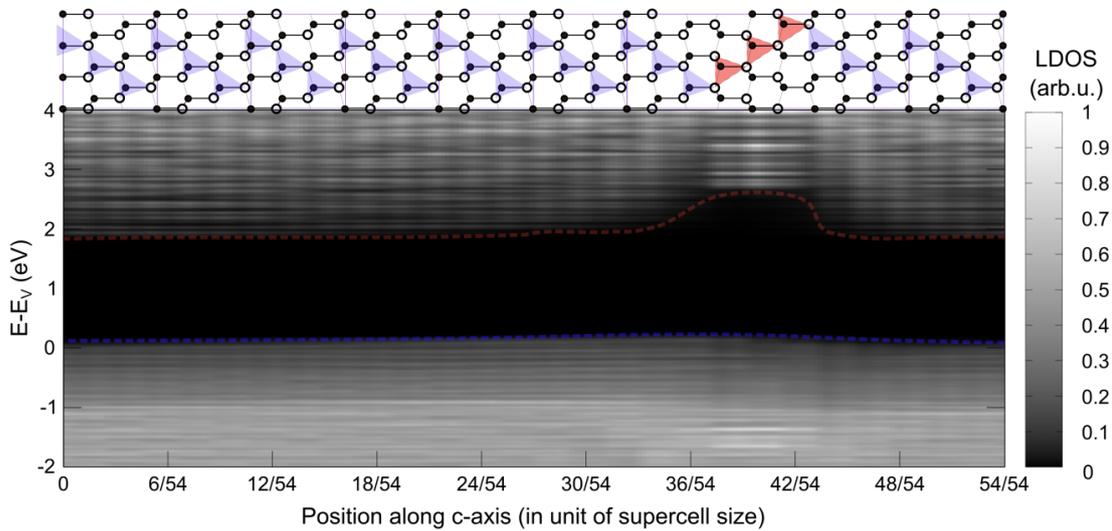

**Fig. 2** Local density of states (LDOS) integrated in slice boxes along the c-axis of the simulated supercell, which is illustrated on top. The red triangles highlight the tetrahedra of the faulted planes as compared to the perfect stacking of the 3C-SiC (blue tetrahedra). The dotted lines highlight the trend of the band edges along the c-axis: a quantum barrier is very evident in the space region where the triple SF is present.

**Terminating single SF by partial dislocations.** Atomistic details of the core dislocations may have a crucial impact on the electronic properties of the material. By exploiting the structures obtained after the MD, we investigated by means of DFT calculations different possible atomistic configurations for the partial dislocations terminating the single, double and triple SFs with two main goals: finding the most thermodynamically stable configurations and among them identifying those detrimental to the electronic properties of the devices. Between all possible configurations, we focus on those that do not have Si or C dangling bonds. In fact, Bernardini et al.[66] found for the (single) 30° partial dislocations in SiC that unreconstructed geometry for both Si- and C-core partials, showing Si or C dangling bonds, are energetically unfavoured, as compared to the reconstructed case. The latter shows Si-Si or C-C covalent bonds, which saturate the dangling bonds of the unreconstructed geometry, pushing the defect states outside the bandgap or close to valence band edge in case of Si-core 30° partial dislocations. In particular, the strong C-C bonds of the C-core of the 30° partial dislocations (configuration on the left in Fig.3) enhance the energy separation between bonding and anti-bonding states associated with the hybridization of their orbitals, resembling the $sp^3$ hybridization in bulk SiC and thus preventing the formation of intra-gap states as evident in Fig. 3 (right). On the contrary, the Si-Si bonds of the Si-core of the 30° partial dislocations, illustrated on the right side of Fig. 3, is much weaker than the C-Si bonds in the pristine SiC. As matter of the fact, the energy separation between bonding and anti-bonding states of the Si-core atoms is not as large as in the pristine SiC case, and intra-gap states are visibly formed close to the VB edge (see Fig. 3). The partial density of states (PDOS) associated with C and Si atoms at the core of the partial dislocation or in the pristine bulk, as plotted in Fig. 4, better clarifies these aspects. The VB of the 3C-SiC is mainly originated by orbitals of the C-atoms, while the Si atoms contribute more to the CB states. When C-C (or Si-Si) bonds are present in SiC, the gap between bonding and anti-bonding states is due to the hybridization between orbitals of the same atomic type, which then contribute both to the formation of the CB and VB. In fact, the PODS of the C-core atoms plotted in Fig. 4 shows a marked peak in the CB, which is not present in the PDOS of C-bulk atoms, but this peak is well above the CB edge. The Si-core atoms induce a PDOS peak in the VB, also not present in the bulk case. This peak evidences states that are above the VB edge, reasonably due to the weaker Si-Si bonds, as compared to the C-Si ones, and forming band-like states from the VB edge up to about 0.7 eV above the VB. The band-like behaviour of these states will be discussed further below.

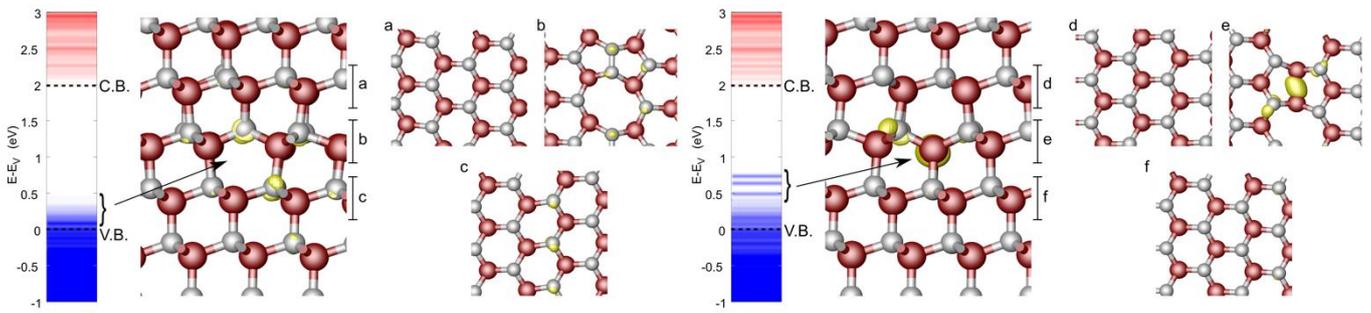

**Fig. 3** Defect states at the C-core (on the left) and at the Si-core (on the right) of the 30° partial dislocation: the Kohn-Sham eigenvalue spectra of the dislocation cores are plotted, with the blue (red) lines indicating occupied (unoccupied) states and the two dashed lines indicating the valence band (VB) and the conduction band (CB) edges in the pristine SiC. The side views of the dislocation cores with the charge densities of the corresponding gap states are also illustrated, including the top views of the (-111) glide planes (with a, b, c and d, e, f labels) and the charge densities of the gap states. Red and grey balls indicate Si and C atoms, respectively.

The electronic properties of the 90° partial dislocations illustrated in Fig. 5 are very similar to the cases of the 30° partial dislocations, and only the Si-core 90° partial dislocations (see left part of Fig.5) show intra-gap states, but not the C-core ones (right side of Fig.5). These intra-gap states generated by the Si-Si bonds of the 90° partial dislocation are even higher in energy and thus closer to the CB, as compared to the states Si-core 30° partial dislocations, because of the longer Si-Si bonds in the 90° partials. The weaker Si-Si bonds are responsible for the smaller gap between these states and the CB states, in agreement with the description drawn above.

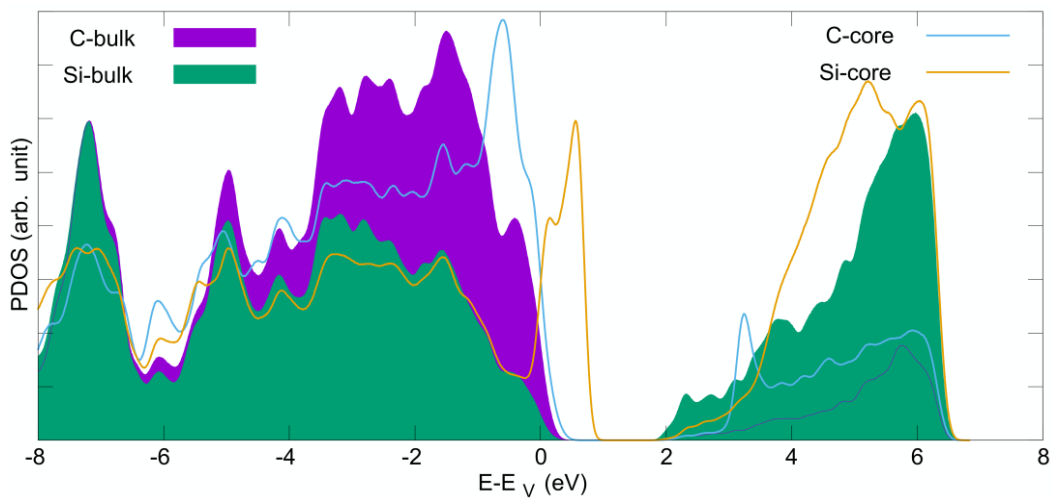

**Fig. 4** Partial density of states (PDOS) of 3C-SiC containing both a Si- and a C-core 30° partial dislocations. The PDOS of C and Si atoms in the core of the dislocation is compared to that of C and Si atoms in the pristine region of the bulk SiC.

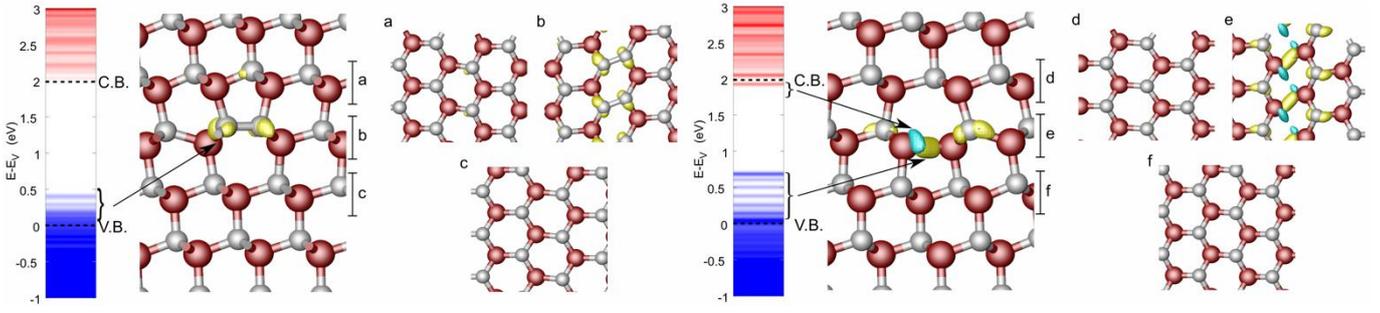

**Fig. 5** Defect states at the C-core (on the left) and at the Si-core (on the right) of the 90° partial dislocation: the Kohn-Sham eigenvalue spectra of the dislocation cores are plotted, with the blue (red) lines indicating occupied (unoccupied) states and the two dashed lines indicating the valence band (VB) and the conduction band (CB) edges in the pristine SiC. The side views of the dislocation cores with the charge densities of the corresponding gap states are also illustrated, including the top views of the (-111) glide planes (with a, b, c and d, e, f labels) and the charge densities of the gap states. Red and grey balls indicate Si and C atoms, respectively. The different colors of the charge densities highlight the different integration energy ranges.

**Double SF.** The results on the electronic properties of the partial dislocations terminating the single SF in 3C-SiC evidence the potential impact of these defects on the electronic properties of 3C-SiC. Still, the intra-gap states due to the Si-core partial dislocations are found at energies about 1.5 eV lower than the CB, minimum, probably too far in energy from the conduction band to be the only cause of trapping-assisted recombination/tunneling processes,[20,64] which have been related to the high leakage current experimentally observed in 3C-SiC devices. However, we have shown in a different work how double and triple SFs are markedly more stable than the single ones.[7] The MD simulations have also shown how the termination of the double and triple SFs are central in the formation mechanisms and evolution of dislocations (see Fig. 1), and they have highlighted the structural complexity of these dislocations, which could induce electronic defects states even worse that those found for the dislocations terminating the single SFs.

We found two main stable structures for the dislocations terminating the double SFs and they are illustrated in Fig.6. They are basically composed of a 30° partial on top of a 90° partial: the structure "A" illustrated on the left side is composed of a C-core 30° partial (see inset a) on top of a Si-core 90° partial (see inset b), while the structure "B", on the right side, shows a Si-core 30° partial (see inset d) on top of a C-core 90° partial (see inset e). The band structure corresponding to the configuration A does not evidence particular intra-gap states, still there are few states that vary similarly to those found for the simple Si-core 90° partial. Indeed, also in this case the Si-Si bond of the 90° partial dislocation causes the deepest defect states, as evidenced in the charge density plot of Fig. 8. On the contrary, nothing is found in the gap as due to the C-C bond of the 30° partial dislocation, in agreement with results found for the dislocations terminating the single SF.

The configuration B does show evident deep defect states, mostly due to the Si-core 30° partial dislocations, and extending much higher in energy as compared to the Si-core dislocations terminating the single SF. Indeed, the charge density plots show that the Si-core 30° partial dislocations are responsible for a dispersive band lying in the SiC bandgap, approximately from 0.7 eV to 1.1 eV above the VB edge. Although the C-core 90° partial dislocation does not contribute directly to these states, as evidenced by the charge density plot, the C atoms of the dislocation core are bonded to the Si atoms of the Si-core 30° partial dislocation lying above. Because of the stronger C-C bonds as compared to the Si-C bonds, the hybridization of the Si-core atoms (which are themselves bonded to another Si atom, two bulk-like C and one C-core atoms) gets even weaker, with the bonding and anti-bonding states having a smaller gap as compared to the single Si-core 30° partial dislocation. This is likely the main reason for the up-shift of the intra-gap states.

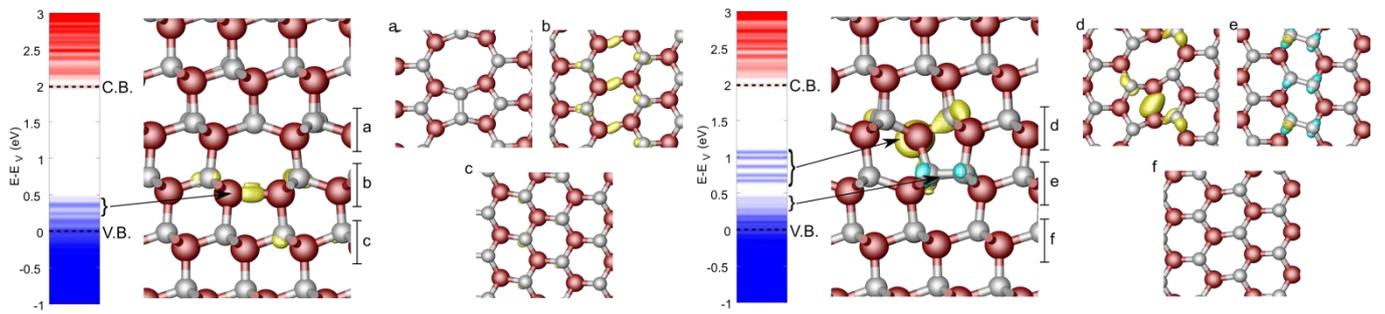

**Fig. 6** Defect states at the double dislocation complexes: configuration A on the left and configuration B on the right. The Kohn-Sham eigenvalue spectra of the dislocation cores are plotted, with the blue (red) lines indicating occupied (unoccupied) states and the two dashed lines indicating VB and CB edges in the pristine SiC. The side views of the dislocation cores are illustrated and the corresponding top views of the (-111) glide planes are shown in the insets a, b, c and d, e, f. Red and gray balls indicate Si and C atoms, respectively. The different colors of the charge densities highlight the different integration energy ranges.

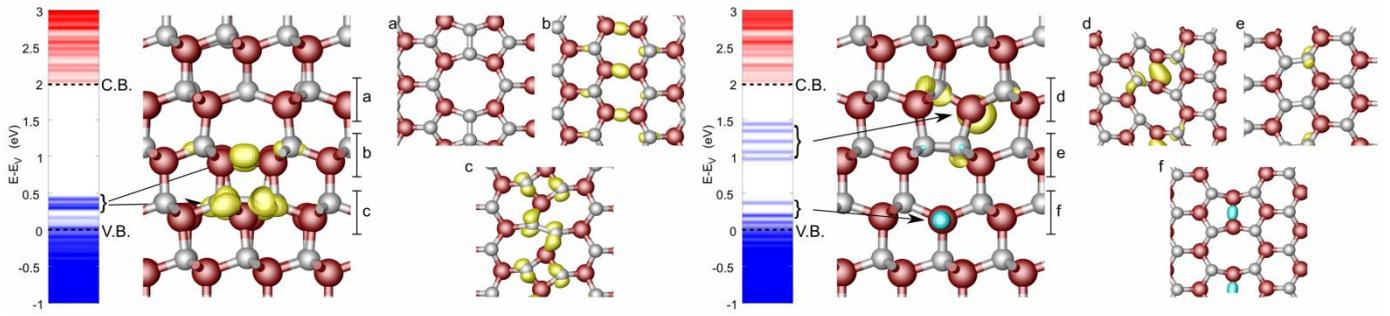

**Fig. 7** Defect states at the triple dislocation complex: configuration A on the left and configuration B on the right. The Kohn-Sham eigenvalue spectra of the dislocation cores are plotted, with the blue (red) lines indicating occupied (unoccupied) states and the two dashed lines indicating VB and CB edges in the pristine SiC. The side views of the dislocation cores with the charge densities of the corresponding gap states are also illustrated, including the top views of the (-111) glide planes (as in the side view with a, b, c and d, e, f labels) and the charge densities of the gap states. Red and gray balls indicate Si and C atoms, respectively. The different colors of the charge densities highlight the different integration energy ranges.

**Triple SF.** For the triple dislocation complex, we investigate some candidate structural configurations after the MD simulations, and we show the two most stable ones in Fig.7. In analogy with the double SF case discussed above, these defects are composed of the stacking of three partial dislocations, two of them (top and bottom ones) are 30° and have the same core species (e.g. C-core for configuration A, on the left, while Si-core for configuration B, on the right), while the intermediate one has 90° character and has the different core species (e.g. Si-core for the configuration A, on the left, C-core for the configuration B, on the right).

In fact, we find that two dislocations of the configuration A (illustrated in the inset a and b of Fig.7) are very similar to the single 30° C- and 90° Si-core partials, thus their defective states are outside the gap (C-core) or close to the valence band-edge (Si-core), very similarly to the configuration A of the double SF discussed above. But this triple SF termination has, in addition, another single 30° C-core partial (see inset (c) of Fig.7), which although showing a slightly different reconstruction than the top one, has defect states laying still in the same energy range (below 0.5 eV from the VB), as originating from the bonds of the C-core atoms.

On the contrary, Fig.7 evidences that the configuration B (shown on the right side) having two 30° Si-core partials and a 90° C-core one in between them, is more problematic in terms of defect energy states. In fact, similar to the double SF case, the (top) 30° Si-core partial is responsible for a dispersive band lying in the SiC bandgap from about 1.0 eV to about 1.5 eV above the VB edge (see inset d). As compared to the termination of the double SF (configuration B), the defect energy band of this defect is almost identical but shifted upward by about 0.4 eV. This shift is mainly due to the different Si-Si bond length of the 30° Si-core partials: the termination of the double SF has a Si-Si bond length of about 2.33 Å as compared to 2.39 Å of the top 30° Si-core partial in the termination of the triple SF. The larger bond length of the 30° Si-core partial in the triple SF, originating from the particular stacking geometry of the three partials, leads to even weaker bonds as compared to the Si-Si bonds in the double termination, thus reducing further the gap between bonding and anti-bonding states. While the top 30° Si-core partial introduces these deep states in the 3C-SiC gap, the bottom one has states closer to the VB (see inset f), and very similar to those of the single 30° Si-core partial dislocation. These is due to the fact that the Si atoms of the top partial are bonded to the C-core atoms of the 90° middle partial, as in the case of the termination of the double SF, while the bottom 30° Si-core partial is in the simple configuration as the single 30° partial, but just having slightly shorter Si-Si bond length (2.36 Å as compared to 2.37 Å of the single one).

Finally, in Fig. 8 the band structure along the <110> direction for the supercell containing the triple SF and the terminations of Fig.7 is plotted. One can clearly see that the 5 energy levels appearing on the right side of Fig.7 (as well as in Fig. 6, right side) and forming the defect energy bands at 1.0-1.5 eV (and 0.6-1.1 eV), are not different energy states, but a single dispersive branch, originating from the periodicity of the defect along the dislocation line. The discrete energy levels in Fig.6 and Fig.7 are then an artifact of the limited K-points employed for the calculation of the LDOS.

Thus, we conclude that the most plausible candidates as killer defects in 3C-SiC are the partials terminating the double and triple SFs illustrated in Fig. 6 and Fig.7, where the particular configuration of the Si-core 30° partials pushes the states of the covalent bonding of the Si-core inside the gap of the pristine SiC. Because of the periodicity of the defect along the dislocation line, these defects states are also dispersive in the k-space, thus forming 0.5 eV wide defect bands in the gap of the pristine SiC, while shallow defects are expected for the termination of the single SF. Note that the triple dislocation complex is structurally identical to the complexes forming the Σ3 {112} incoherent boundaries likely present in more intricate defects, such as grain boundaries, which are occasionally found in the epilayers. Therefore, such defects may lead to very similar electronic defect states.[59,60,61,65]

The coexistence of double and triple SFs in 3C-SiC (with a higher population of the latter), together with a minor occurrence of the single SFs is clearly shown in early experiments.[39] Therefore, our results suggest that this whole usual defect population is responsible for intra-gap states filling the gap of 3C-SiC up to about 1.5 eV from the VB maximum.

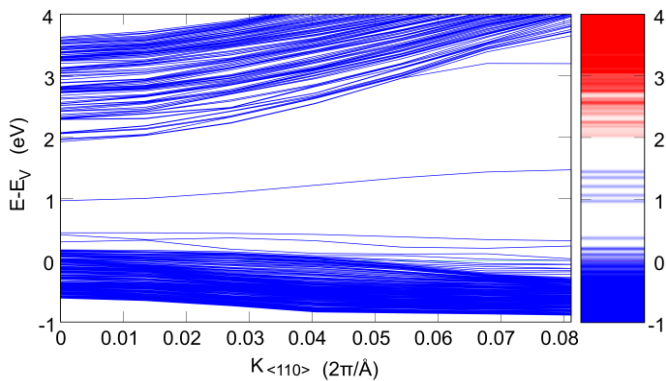

**Fig. 8** Band structure along the <110> direction for the supercell containing the terminations of the triple SF. The deep defect states in the LDOS are clearly due to a single dispersive band.

## Conclusions

We have presented an in-depth theoretical/computational analysis of the most typical extended defects in 3C-SiC/Si(001). Large-scale MD simulations based on the classical Vashista potential[41] were exploited to investigate the physical origin of defect complexes typically observed in experiments. In particular, we have shown that partial dislocation complexes terminating triple SFs can be easily generated during dislocation dynamics, under the compressive strain conditions which follow the formation of sessile dislocation arrays at the 3C-SiC/Si interface. Configurations revealed by MD simulations were subsequently analysed and re-converged by DFT calculations. This multiscale approach was precious in saving computational time: by using DFT only, the computational cost would have precluded the exploration of defect dynamics on any meaningful time scale.

After completing the structural analysis, we have investigated the influence of the defects on the electronic properties of 3C-SiC. Meta-GGA exchange-correlation functionals have been exploited in order to improve the bandgap prediction. It turned out that the net result of a distribution of defects including single, double, and triple SF is a significant filling of the gap, with detrimental consequences in terms of leakage currents. As defect states induced by the partial dislocation terminating the single SF are observed to lay close to the edge of the valence band, we conclude that dislocation complexes, terminating double and triple SF, are killer defects in C-SiC.

Based on these findings, a next important step in the development of high-quality epitaxial 3C-SiC would involve lowering the concentration of partial dislocation complexes. Further experimental and theoretical research is surely needed in order to find optimal growth conditions for achieving such a result.

We notice that the solution of raising the temperature to lower the defect density, typically adopted in the epitaxy for several different systems, might not work in this case. In a recent paper indeed, we showed that at high temperatures thermodynamics favours hexagonal SiC vs 3C-SiC.[7] As multiple SFs correspond to insertion of hexagonal phases within the cubic one, raising temperatures too much might be counterproductive. As an alternative, to reducing the effect of dislocation complexes, one could aim at tailoring their core, and push the intra-gap defect states out of the gap. Controlled insertion of contaminants could help in this respect.

## Acknowledgements

Authors acknowledge EU for funding the CHALLENGE project (3C-SiC Hetero-epitaxiALLy grown on silicon compliancE substrates and 3C-SiC substrates for sustaiNable wide-band-Gap power devices) within the EU's H2020 framework program for research and innovation under grant agreement n. 720827. The CINECA award under the ISCRA initiative, for the availability of high-performance computing resources and support, is also acknowledged.